# Temporal optical neurons for serial deep learning

**Authors:**

Zhixing Lin,[1,2,3], Shuqian Sun [1,2,3], José Azaña, [4] Wei Li [1,2,3], Ninghua Zhu[1,2,3], and Ming Li,[1,2,3,*]

**Affiliations:**

[1] State Key Laboratory on Integrated Optoelectronics, Institute of Semiconductors, Chinese Academy of Sciences, Beijing 100083, China

[2] School of Electronic, Electrical and Communication Engineering, University of Chinese Academy of Sciences, Beijing 100049, China

[3] Center of Materials Science and Optoelectronics Engineering, University of Chinese Academy of Sciences, Beijing 100190, China

[4] Institut National de la Recherche Scientifique – Centre Energie, Matériaux et Télécommunications (INRS-EMT), Montreal, Quebec, Canada.

**E-mail:**

*e-mail: ml@semi.ac.cn

**Abstract:**

**Deep learning is able to functionally simulate the human brain and thus, it has attracted considerable interest. Optics-assisted deep learning is a promising approach to improve the forward-propagation speed and reduce the power consumption. However, present methods are based on a parallel processing approach that is inherently ineffective in dealing with serial data signals at the core of information and communication technologies. Here, we propose and demonstrate a serial optical deep learning concept that is specifically designed to directly process high-speed temporal data. By utilizing ultra-short coherent optical pulses as the information carriers, the neurons are distributed at different time slots in a serial pattern, and interconnected to each other through group delay dispersion. A 4-layer serial optical neural network (SONN) was constructed and trained for classification of both analog and digital signals with simulated accuracy rates of over 90% with proper individuality variance rates. Furthermore, we performed a proof-of-concept experiment of a pseudo-3-layer SONN to successfully recognize the ASCII (American Standard Code for Information Interchange) codes of English letters at a data rate of 12 Gbps. This concept represents a novel one-dimensional realization of artificial neural networks, enabling an efficient application of optical deep learning methods to the analysis and processing of serial data signals, while offering a new overall perspective for the temporal signal processing.**

# 1. Introduction

The framework of artificial neural networks is originated from biological neural networks and has been widely used to implement deep learning. The realization of deep learning requires a tremendous amount of data and computational resources. Fortunately, the advent of the big data field and the explosion of computing capabilities, supported by the deployment of graphics processing units (GPUs) and other parallel processing units, have enabled a remarkable progress in deep learning. As a result, deep learning has been extensively employed for many different important tasks both in industrial and academic settings, such as for speech recognition, image classification, game playing (decision making), language translation, etc.[1–7]. However, computing speed and power consumption aspects remain as key concerns for further development of deep learning methods based on conventional technologies.

It stands to reason that the nature of light endows photonic signal processing with broad bandwidth (i.e., potential high processing speed), low delay/latency, low power consumption, and advanced mathematical operation capabilities. Therefore, several different approaches have been proposed and demonstrated for the implementation of neural networks using light waves, so-called optical neural networks (ONNs), including deep learning through optical diffraction, optical interference, multimode fiber, wavelength division multiplexing, etc.[8–11]. As shown in Fig. 1a, all these schemes operate on parallel-pattern data in the physical space domain, which are processed through a system that emulates a traditional neural network architecture. This parallel-signal-processing structure can only be fed with data in a parallel fashion. Nevertheless, serial data are utilized across a broad range of application scenarios, such as in telecommunications, sensing, lidar, radar, ultra-fast optical imaging etc.[12–19] Thus, the data present in these applications are inherently incompatible with current ONNs. Although serial-to-parallel conversion may be envisioned, this would lead to increased processing latencies and reduced overall system efficiency, limiting the possibility of application of ONN strategies on high-speed serial data signals.

In this paper, we propose an entirely new ONN scheme that is specifically conceived for processing a serial data flow. In this scheme, the serial pattern to be processed (the object) is encoded in a sequence of optical pulses, which serves as a sequential set of temporal neurons that are subsequently interconnected through group-velocity dispersion, e.g., easily implemented by linear propagation through a section of optical fiber. In this way, the optical neurons are distributed and interconnected with each other along a single channel, namely, the time domain. The desired 'weights' on the different neurons are realized by imposing a prescribed modulation on the coded pulses along the time domain, e.g., through widely available temporal modulation. This time-domain serial ONN (SONN) scheme is thus ideally suited to deal with serial data patterns, directly on-the-fly and in a real-time fashion. We discuss here the basic design conditions and main trade-offs of the proposed SOON scheme. Moreover, the concept is successfully validated for analog and digital data by numerical simulations and a proof-of-concept experiment.

## 2. Principle

Figure 1a shows the schematic of a conventional optical network for deep learning, in which the neurons are arranged abreast, and the 'data' flows through the network in a parallel fashion. The proposed new SONN scheme is illustrated in Fig. 1b. The input of this SONN is a continuous sequence of short optical pulses modulated by the input pattern (or object) under analysis. In other words, the data to be recognized is allocated in different timeslots in a serial optical channel, rather than in different parallel channels. For simplicity, in our following analysis, we assume that the object is a temporal intensity (or amplitude) pattern, thus involving temporal intensity modulation of the optical pulse train, though generally, the object can be a complex-field pattern. Subsequently, the modulated pulse sequence is fed into the SONN and serially processed in real-time. In particular, different consecutive time slots in the temporal trace represent the different serial neurons in the network, with each neuron occupying a single period of the original pulse train.

In the SONN shown in Fig. 1c, the neurons of consecutive layers are connected using group-velocity dispersion. As it is well known, dispersion retards or advances the different spectral components of a propagating signal to different time slots. As shown in Fig. 1c, when the pulse train propagates through a dispersive medium, the broadband spectral components of each pulse (arrows in different colors) is distributed to adjacent timeslots, and add coherently. This process is used to perform the required connections among the neurons in consecutive layers. The amount of dispersion used in the scheme should be sufficiently high. By doing so, the spectral content of each pulse can be spread out in time beyond the number of periods (or neurons) to be connected. Within these fundamental condition, our numerical simulations show that the performance of the proposed SOON scheme is optimized when the dispersion amount is fixed to work under the required conditions to produce a Talbot self-imaging effect of the original pulse train[20–22] (see Supplementary Section 4 for a detailed discussion).

Recall that temporal Talbot self-imaging can be observed when a periodic pulse train propagates through a dispersive medium that satisfies a well-defined condition, namely, $s/f_{rep}^2$ = $2\pi|\ddot{\beta}|L_T$, where $f_{rep}$ is the repetition rate of the input pulse train, and $|\ddot{\beta}|$, $L_T$ and s are the second-order derivative of the propagation constant, the dispersive propagation length and a positive integer number, respectively. At the dispersion lengths defined by this condition, the original periodic pulse train reproduces itself with the exact same repetition rate. As per our discussions in Supplementary Section 4, operating at a Talbot distance ensures an optimal interaction among the consecutive coded pulses (neurons) to be interconnected. Thus, the dispersion value of each layer in the network should be carefully designed to satisfy a Talbot effect condition[20].

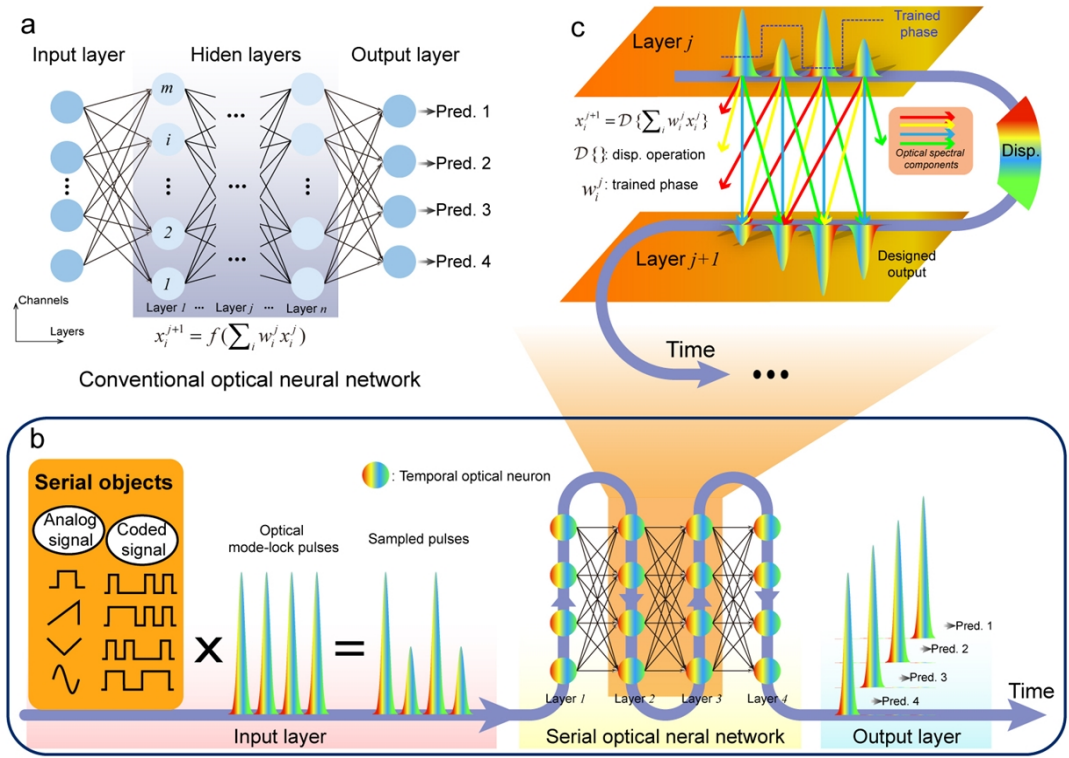

**Fig. 1| Principle of SONN. a**, Schematic of a conventional optical network for deep learning. The neurons are arranged abreast, and the 'data' flows through the network in a parallel fashion. **b**, Schematic of the proposed serial optical neural network (SONN). The objects to be processed are temporal waveforms, which are sampled in amplitude by an optical pulse train (with a constant phase) and then fed into the SONN. All the neurons/pulses are distributed within a dispersive link (e.g., an optical fiber) over time and interconnected through dispersion. **c**, Schematic of the interconnections between layers. The "weights" are applied across the neurons at each layer through a process of temporal complex-field (e.g., phase) modulation of the incoming temporal sequence. The blue dash lines represent the trained phase applied to the neurons/pulses. The optical spectral components (arrows with different colors) of each pulse/neuron in layer $j$ will be retarded or advanced with respect to each other, and then added coherently with the components of other neurons to form the new neuronal pattern in layer $j+1$.

The interconnection of the temporal neurons in between neighboring layers merely through the use of dispersive propagation is not enough to implement a deep learning process because of the lack of the 'weights', the essential factor of any neural network scheme. In our proposed system, the 'weights' are realized by imposing a proper temporal modulation pattern on the sequential pulses/neurons before propagating the resulting sequence through the dispersive medium. Phase modulation is used in our demonstrations reported here, though complex modulation (both amplitude and phase) is generally possible, as discussed further in the Supplementary Section 7. The resulting complex-valued 'weights' of the neurons are then determined by the temporal modulation function—to be customized through the training stage—and the phase imposed by the dispersive propagation process itself. As stated before, the connections among neurons are produced through coherent addition of the temporally dispersed optical spectral components of the phase-modulated neurons of the former layer, as shown in Fig. 1c. To be more specific, each of the spectrum components (colorful arrows shown in Fig. 1c) in layer $j$ are relocated to adjacent timeslots through dispersive propagation at a Talbot distance $L_T$. Then they add coherently with respective weights that depend on the trained phase applied to each neuron. At the subsequent network layer, they form new neurons. Without phase modulation, the dispersion alone will only produce a temporal averaging effect to the original 'intensity-varying' pulses[23]. However, in our proposed scheme, the output of each layer can be controlled through the application of the trained phase profile to the temporal neurons. By properly training the phase modulation profiles to be applied across all the layers in the network, one can customize the shape of the temporal waveform at the output of the network according to the specific features of the input data pattern (object) under analysis, implementing a desired input-to-output waveform mapping. For instance, as illustrated in Fig. 1b, a final output waveform with a peak pulse at a designed temporal position can be obtained, with the temporal position being dependent on the specific shape of the input object. In this way, different input data patterns can be recognized through the different time positions of the peak pulses at the ONN output. Hence, through this new ONN concept, the two-dimensional structure of a traditional neural network is actually implemented in a single dimension (the temporal dimension), enabling a serial ONN architecture.

Figure 2a shows the model flowsheet of the proposed SONN and related functions that are critical to construct the model. In the forward propagation, every layer contains the processes of temporal phase modulation and dispersive propagation. After propagation through several layers, the final measured output waveform together with the target (ideal) output waveform are used to calculate the corresponding mean square error (MSE) as the cost function. A backpropagation algorithm is employed to train the phase modulation profiles across all layers of the network to achieve the highest possible accuracy rate (see Supplementary Section 1 for a detailed analysis of this SONN model). In what follows, we illustrate the proposed concept through a 4-layer serial neural network design for the classifications of analog and digital signals, respectively.

## 3. Simulation

An example of a 4-layer SONN based on the proposed architecture is first designed and validated by numerical simulations. The simulation results for the classification of analog signals are shown in Fig. 2b-m. The data for training and testing were specifically designed to validate the SONN architecture and evaluate its performance. A pulse train with a repetition rate of 5 GHz is used to sample the different objects under analysis, namely, sine, square, reverse triangle and sawtooth waves, each repeating periodically with a 3-ns period. A 30-dB individuality variance rate (IVR) on peak amplitude was also imposed on the patterns/objects to diversify the data, with the aim of generalizing the trained model to unseen data (see Supplementary Section 6 for more information about IVR). The "features" data (input objects to be processed) are shown in Fig. 2b—e and the "label" data (corresponding ideal target output waveforms) are shown in Fig. 2f—i. We generated 100 data for each kind of analog wave (400 data in total), of which 70% were used to train the network and 30% were used to test the training performance. The classification result is correct when the peak pulse of the output waveform matches that of its "label" waveform in terms of temporal position. Given that the data rate of the phase profile is 10 Gb/s and the batch size is 6, the network was trained for 1000 epochs till the trend of cost and accuracy were stable (see Supplementary Section 2). Noticeably, the accuracy can reach 100% at an IVR of 30 dB (see Supplementary Section 6 for more results). Examples of the output waveforms corresponding to some of the test data are shown in Fig. 2j—m. As can be seen, the peak pulses of the trained outputs align well with those of the 'label' waveforms.

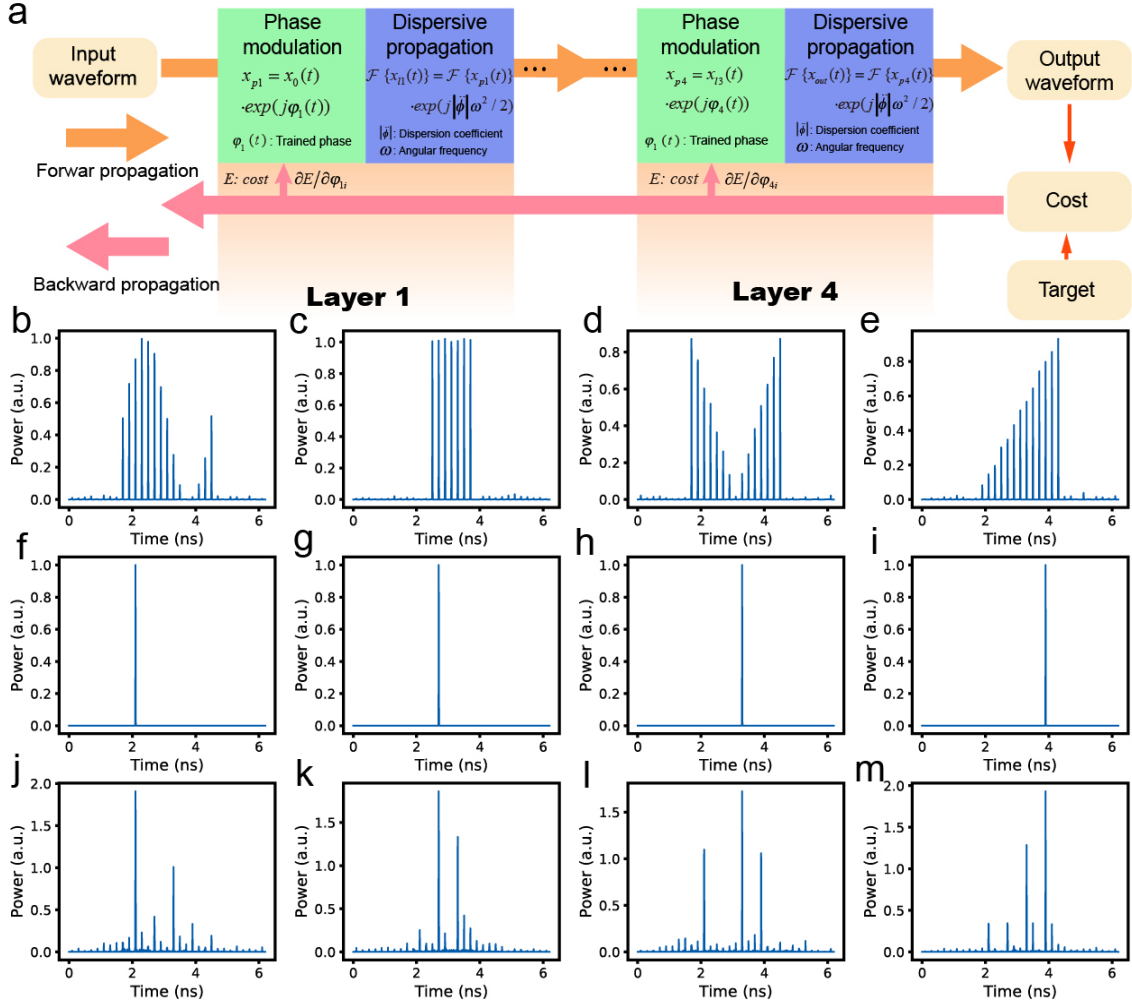

**Fig. 2| Simulation of a 4-layer SONN for analog signals. a**, Model of the SONN consisting of forward and backward propagation. Each layer contains two operations — temporal phase modulation and dispersive propagation. The related mathematical functions are shown in the corresponding regions. Layer 2&3 are omitted considering the similar structure of all layers. **b—e**, 'Feature' waveforms of sine, square, reverse-triangle and sawtooth functions, respectively. **f—i,** Ideal 'label' waveforms corresponding to **b—e**. **j—m,** Trained output waveforms corresponding to **b—e**.

The operation of the proposed SONN is based on the interference among the optical spectral components of the incoming pulses. These components cancel each other at the specific time slots for which no pulse is desired, as shown in Fig. 2f—i. Nonetheless, in the resulting output waveforms, we observe the presence of undesired side peaks near the main pulse. These peaks have no negative effect on the classification accuracy. Another important issue to be considered for this optical deep learning scheme concerns the classification of successive patterns/objects. Due to the broadband nature of the light source and the dispersion used in the SONN, the optical spectral components of an incoming pattern will be inevitably stretched to beyond the temporal slot allocated to this specific pattern, which may affect the classification ability for consecutive patterns/objects. However, by inserting suitable temporal gaps between the incoming patterns and employing a symmetrical dispersion strategy, consecutive patterns can be classified with the desired accuracy, as shown in Supplementary Section 5. Symmetrical dispersion here refers to the use of an amount of dispersion in neighboring layers with the same absolute value but with alternating (opposite) signs. This strategy also allows one to overcome a potential limitation in the number of layers associated with an excessive accumulation of dispersion along the system, thus helping to implement a 'deeper' ONN. Here, the gap in the evaluated models for classification of analog signals (results in Fig. 2) and digital signals (results in Fig. 3, described below) is fixed to be 16× and 20× the pulse repetition period, and the absolute dispersion values for these two cases are $2 \times D_T$ and $3 \times D_T$, respectively, where $D_T = 1/2\pi f_{rep}^2$ is the fundamental (first-order) Talbot length, i.e., with s = 1, corresponding to the input pulse train at $f_{rep} =$ 5 GHz. The additional latency caused by the time gap should be taken into consideration in evaluating the total delay needed for the analysis of each pattern. In the considered examples, an additional latency of ~3.2 ns and ~4 ns should be considered for the analog and digital patterns under analysis, respectively.

Next, the classification capability for digital signals is investigated. The digital signals to be recognized are 8-bit binary ASCII (American Standard Code for Information Interchange) codes— 'u', 'c', 'a' and 's'. These codes are transferred to the pulses, as shown in Fig. 3a—d through on-off-keying (OOK) intensity modulation. The corresponding "label" waveforms are presented in Fig. 3e—h. The model and the hyperparameters are the same as those in the classification model for analog signals. An accuracy of 100% was obtained after an 800-epoch training at an IVR of 30 dB (see Supplementary Section 6 for more results with different IVRs). The relationship between IVR and accuracy rate is discussed in the Supplementary Section 6. The outputs of the test data after training are displayed in Fig. 3i—l, where the peak position also matches well with the corresponding "label" waveform in Fig. 3e—h.

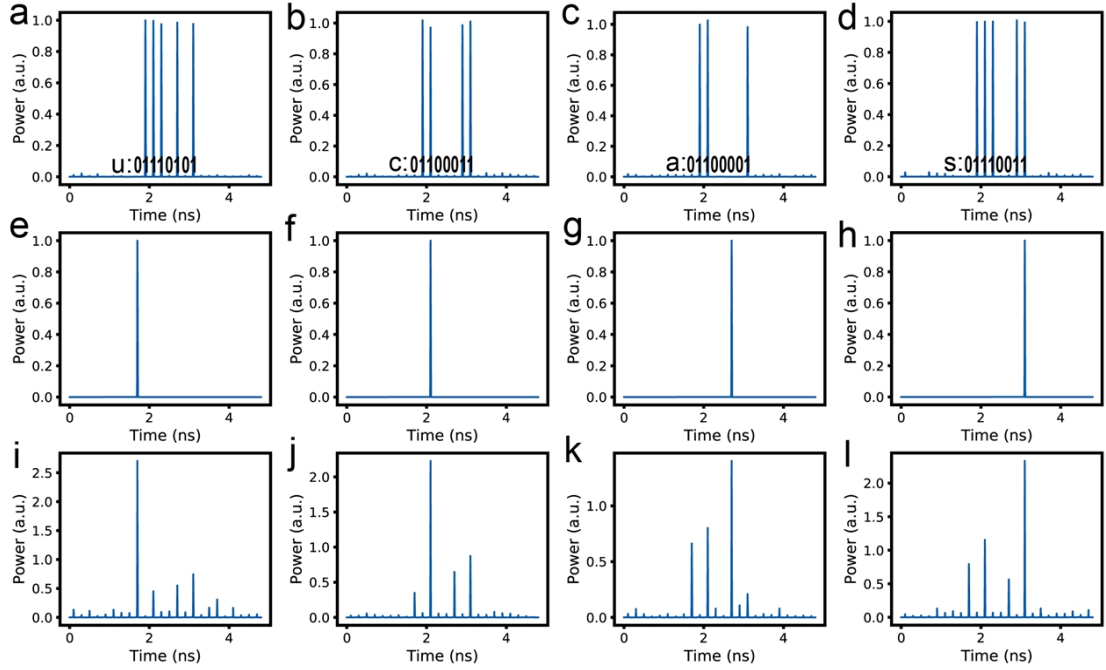

**Fig. 3| Simulation of a 4-layer SONN for digital signals. a—d**, 'Feature' waveforms of four English letters — 'u', 'c', 'a' and 's'— in the form of a binary ASCII code. Each bit is represented by the power of each pulse. **e—h**, Ideal 'label' waveforms corresponding to **a—d**. **i—l**, Trained output waveforms corresponding to **a—d**.

## 4. Experiment

Proof-of-concept experiments were also performed in order to verify the feasibility of the proposed SONN model. We designed and trained a pseudo-3-layer SONN for the separate classifications of two groups of letters— 'u' versus 'c' and 'a' versus 's'. The experimental setup is presented in Fig. 4a (see Methods for more details). Due to the degradation in the recognition capability as compared with a 4-layer SONN, a layer with no trainable variable (no phase modulation) is added for an improved performance (see Supplementary Section 3). The repetition rate of the mode-locked laser (MLL) is set to be 12 GHz, matched with the dispersion value (~ −850 ps nm$^{-1}$ for the first-order Talbot condition). The ideal output waveform (not shown here) is set to be a peak pulse located on the left or right side to better discriminate the results. We trained the pseudo-3-layer ONN with 2000 and 1000 epochs and obtained the phase modulation profiles which are shown inside the yellow arrows in Fig. 4a (red line for 'u' & 'c', blue line for 'a' & 's'). Notice that unlike the symmetrical dispersion strategy used in the classification setup simulated above, the dispersion of each layer in the experimental design is fixed to $D_T$ with the same sign. The data rate of the phase modulation profiles is also reduced with respect to the design above, and in particular, it is fixed to be equal to the data rate of the input data pattern. These two factors give rise to an amount of undesired noise around the peak pulse in the obtained output waveforms, affecting the classification performance of the demonstrated SONN, as detailed below.

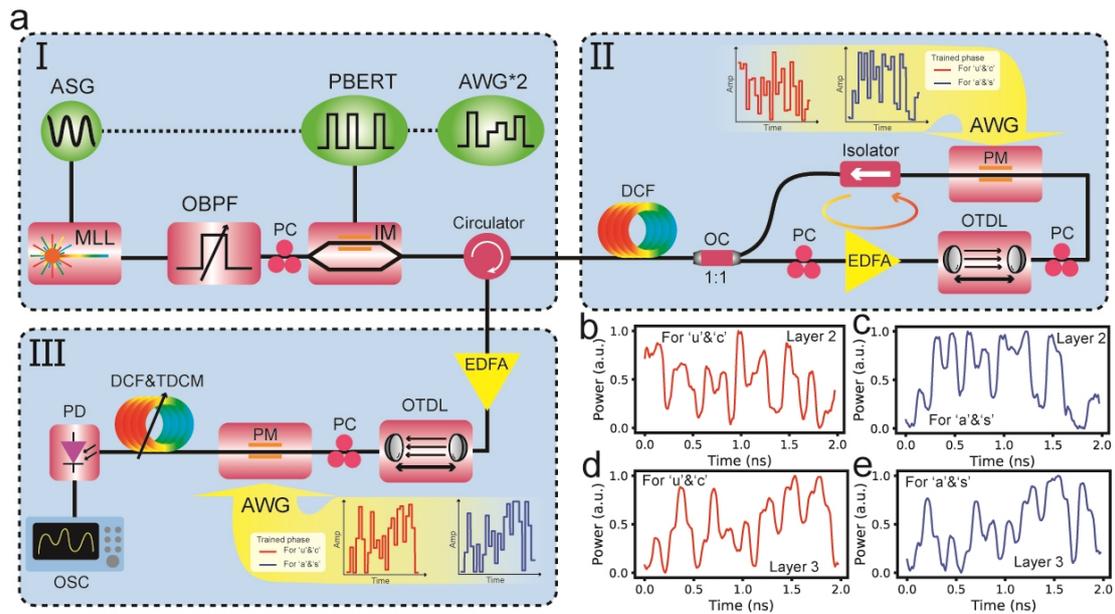

**Fig. 4| Experimental setup of a pseudo-3-layer SONN. a**, Experimental configuration of a pseudo-3-layer SONN. Ⅰ: Generation and coding of the pulse train. ASG, analog signal generation; MLL, actively mode-locked laser; OBPF, optical bandpass filter; PC, polarization controller; PBERT, parallel bit error ratio tester; IM, intensity modulator; AWG, arbitrary waveform generator; Ⅱ: Pseudo layer 1 and layer 2 with phase modulation of the proposed SONN. DCF, dispersion compensation fiber; OC: optical coupler; EDFA, erbium-doped fiber amplifier; OTDL, optical tunable delay line; PM, phase modulator; Pseudo layer 1 contains no trainable variables, i.e., modulation phase, and it is added in the system in order to obtain a better classification performance (see supplementary information). Ⅲ: Layer 3 of the proposed SONN. TDCM, tunable dispersion compensation module; PD, photodetector; OSC, oscilloscope. The temporal waveforms inside the yellow arrows in Ⅱ and Ⅲ are applied to the phase modulators. Dash lines stand for synchronization between different signal sources. **b**, Trained phase profiles applied on the phase modulators in experiments. The red and blue waveforms are for the 'u' & 'c' and 'a' & 's' classifications, respectively. The waveforms of the first and second row are generated for layer 2 and 3, respectively.

Figures 5a and 5b show the input waveforms corresponding to the English letters of 'u' & 'c' and 'a' & 's', respectively. After carefully synchronizing the optical pulses with the applied trained phase (shown in Fig. 4b, the red line for 'u' & 'c' and the blue line for 'a' & 's'), the two pairs of English letters are well classified by the SONN. The results are shown in Figs. 5c and 5d where the peak pulses appear in the specific prescribed time positions in both the simulation and experimental results. The mismatch between the simulation and the measurement results in regards to the noise observed around the peak pulse is attributed to the instability of the laser and electrical signal source, as well as the unavoidable noise induced by environmental fluctuations along the fiber devices. This phenomenon has been verified in our simulation by changing the IVR of the input signals. Nevertheless, the noise of incoming signals does not affect the temporal position of the peak pulse, which demonstrates the classification ability of the proposed method under practical conditions. Furthermore, the reconfigurability of the proposed system is confirmed by the classifications of two separate groups of objects. By simply changing the phase profile applied to the phase modulators (see respective phase profiles in Fig. 4b), we can fulfill different tasks by using the same hardware setup. In all evaluated cases, the experimental results agree well with the simulation ones, thus confirming the feasibility of the proposed serial optical deep learning technique.

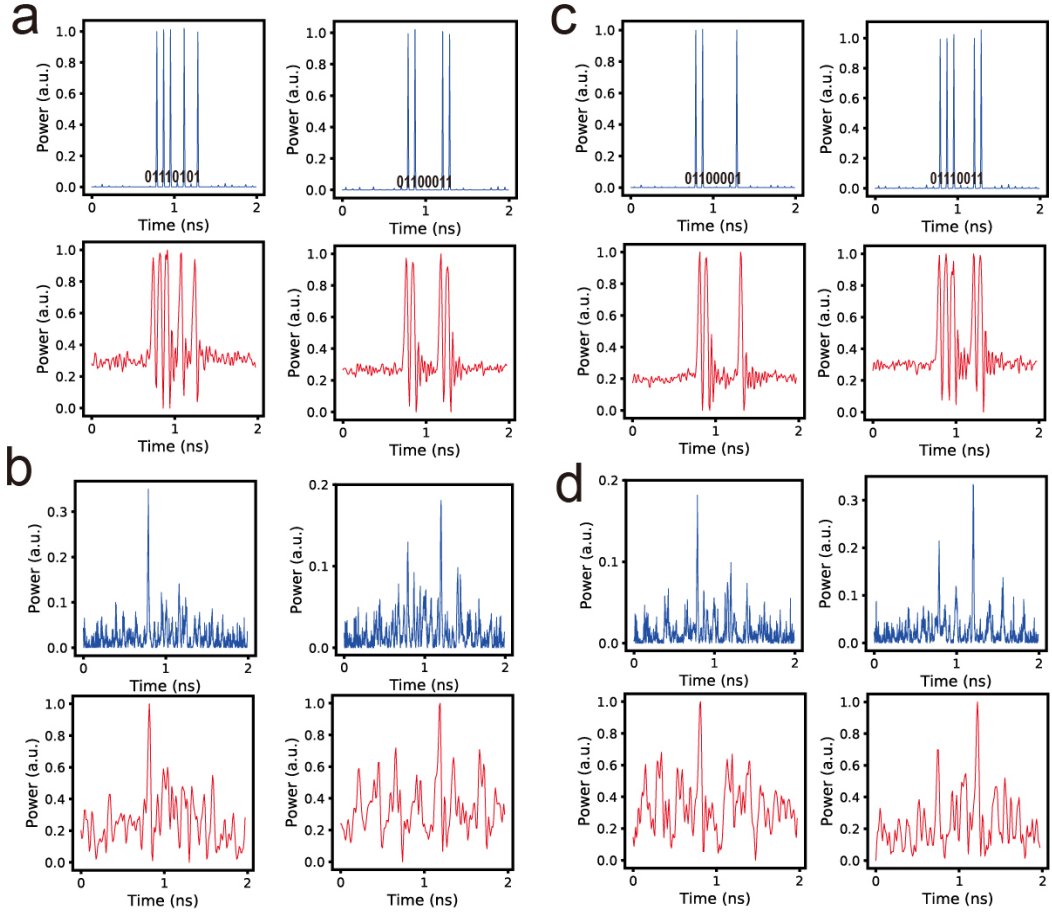

**Fig. 5| Comparison between experimental and simulated results of a pseudo-3-layer SONN. a**, 'Feature' waveforms corresponding to 'u' and 'c' of the input layer in the simulation (blue, top plots) and experiment (red, bottom plots). **b**, Final output waveform corresponding to **a** in the simulation (blue, top plots)/experiment (red, bottom plots). **c**, 'Feature' waveforms corresponding to 'a' and 's' of the input layer in the simulation (blue, top plots)/experiment (red, bottom plots). **d**, Final output waveform corresponding to **c** in the simulation (blue, top plots)/experiment (red, bottom plots). These results were obtained by inputting a single isolated pattern to the SONN at a time.

## 5. Discussion

It is a inherent challenge for traditional ONNs to implement nonlinear 'activation' functions for every neuron because of the side-by-side arrangement of neurons.[8-11] In other words, each neuron in every layer should be provided with a separate nonlinear activation device, which could increase complexity. In contrast, the proposed SONN is ideally suited to the realization of activation functions as all neurons pass through the same single optical fiber. Thus, only one nonlinear activation device would be required for each neural network layer. In practice, the activation functions of this system could be achieved by merely putting one saturable absorber or semiconductor optical amplifier after each layer. A better performance of the SONN could be expected by the incorporation of activation functions[24–26]. In addition, the symmetrical dispersion strategy should facilitate the practical realization of this proposed SONN scheme, e.g., through the use of linearly chirped fiber Bragg gratings (LCFBGs) for implementation of the dispersive media in the network. This is so because a single LCFBG can provide a specific dispersion from one port of the fiber grating, as well as the same amount of dispersion but with the opposite sign from the other port. Thus, a single LCFBG could be used to implement the needed dispersion lines in two neighboring layers, with the grating respectively operated from its two different ports. This strategy would also ensure a higher dispersion precision than using two separate dispersive media in each two consecutive network layers. Furthermore, the use of an LCFBG as the dispersive medium can also reduce the signal processing time (i.e., latency) significantly, thanks to the shorter physical length of the grating as compared with other alternative dispersion devices (e.g., a section of optical fiber). Additionally, it is important to note that the proposed SONN system requires a very strict synchronization between the input pulse train and the electric phase signals, as well as for the output pattern identification. Precise temporal alignment techniques up to the attosecond regime[27,28] are currently available for application on the proposed scheme. Moreover, in addition to the application as a classifier for analog or digital signals demonstrated here, the introduced SONN scheme also holds the potential for implementation of any signal-processing functionality based on a prescribed input-to-output mapping, such as for reconfigurable pulse repetition rate conversion methods, specific waveform extraction, etc.

## 6. Conclusions

In this paper, we have proposed and demonstrated an SONN concept that can process serial data in a direct and real-time fashion without needing the serial-to-parallel conversion based on temporal optical neurons. The concept exploits a novel deep learning strategy that exploits sequential temporal optical neurons, which are suitably weighted through temporal phase modulation and interconnected through group-velocity dispersion. We have provided the design conditions and trade-offs of this novel ONN scheme, and have successfully demonstrated recognition of analog and digital serial data signals within one temporal channel. The proposed SONN is compatible with most current time-domain data processing systems and the data are directly processed in an on-the-fly and real-time manner. By adequately training the phase profiles of the neurons, the 'features' —input data patterns— are well mapped into the desired prescribed 'labels' —output waveforms. We have reported a proof-of-concept experimental demonstration of a pseudo-3-layer SONN, which can recognize the ASCII codes of English letters at a data rate of 12 gigabits per second. Furthermore, the proposed SONN scheme is fully reconfigurable for different tasks and can be readily extended for a simple realization of activation functions. The proposed technique provides a novel route to optical neural networks, potentially enabling to overcome critical speed and efficiency trade-offs of present approaches, and it also offers a fresh overall perspective for temporal signal processing.

**METHODS**

**Training model details.** We constructed the SONN models with Python and Tensorflow-GPU on a computer with two 12-core CPUs and a 2080Ti GPU, and spent approximately 5.5 minutes on training for the classification of analog and digital signals every 1000 epochs. The batch size is 6. The number of neurons at each layer is fixed by the number of phase levels within each period of the temporal phase modulation profile. In this paper, the duration of each phase level is set to half the repetition period of the mode-locked laser, leading to a number of neurons equal to 62 and 48 for the SONNs used for analog signal classification and digital signal classification, respectively. The SONN model with four layers is designed according to the analysis given in the Supplementary Section 1. It is crucial to rearrange the Fourier transform (FT) operations to ensure that the zero-frequency component is shifted to the center of the obtained spectral pattern, i.e., to implement what is usually referred to as an 'fftshift', to directly match the profile of the spectral transfer function of the dispersive media, $H(\omega) = \exp{(j|\ddot{\Phi}|\omega^2/2)}$. However, no function in the Tensorflow module could support this operation. Fortunately, an alternative method operating in the time domain, i.e., inverting every other sampling value of the field function before the FT operation, was used here to mimic the fftshift function. The number of FT points is 210 for every pulse period. Eight periods of zero-amplitude were introduced on each side of the signal under analysis to be sampled by the optical pulse train. The dispersions used in the simulated SOONs were $2D_T, -2D_T, 2D_T, -2D_T$ for the analog signal classification case, and $3D_T, -3D_T, 3D_T, -3D_T$ for the digital signal classification problem ($D_T$ is the dispersion value that satisfies the condition of the first-order temporal Talbot effect). We exploited an Adam optimization algorithm to minimize the loss function, thus maximizing the prediction rate. To obtain an initial point with which the model can achieve a higher accuracy rate, several rounds of training were performed to find the best starting point.

**Experiment.** We demonstrated a pseudo-3-layer SONN rather than a four-layer one due to laboratory limitations (in regards to the number of channels of the electrical signal generators available in our lab). The optical pulse train was generated by an actively mode-locked laser with a clock provided by an analog signal generator (ASG, Keysight, N5183B), and was filtered out-of-cavity by an optical bandpass filter (OBPF, Yenista, XTA-50/W) with a wavelength bandwidth of ~1 nm in order to match the desired specifications for the optical pulses in the SONN. Although the ASG, the arbitrary waveform generators (AWGs, Tektronix, AWG70001A) and the parallel bit error ratio tester (PBERT, Agilent, 81250) were synchronized through a 10-MHz reference signal emitted by the ASG, this synchronization was insufficient and in particular, the relative movement between the signals generated from these instruments could be visualized in a real-time oscilloscope (OSC, Tektronix, DPO73304D). To mitigate the effects of this limited synchronization, we firstly fixed the sampling rate of the AWGs to be 24.112199996 GS s$^{-1}$ (corresponding to two samples per one phase step). Subsequently, through observation of the relative movement on the real-time OSC, we finely tuned the frequencies of the ASG and the PBERT. Finally, the three signals got synchronized when the frequencies of the ASG and the PBERT were set to 12.056099998950 GHz. This procedure helped to reduce significantly the timing jitter among these generated signals, through there was still some remaining jitter even after application of the procedure. As for the alignment between the phase modulation signals and the event-carrying pulses, an optical tunable delay line (OTDL) and several fixed delay lines (each introducing a delay several times longer than the maximal tuning span of the OTDL) were used. The relative position of the phase profile and the pulses in the second layer did not affect the output waveforms, and this only had an effect on the relative position of the resulting waveform over time. Therefore, our efforts focused on aligning the pulses with the phase modulation profile in the third layer only. Besides, the two phase-modulators (EOSpace) employed in this experiment have the same $V_\pi$ of 4.5 V. The phase modulation signals were amplified by broadband low-noise electric amplifiers (SHF, L806A and SHF, S807C) to an appropriate $V_{pp}$. The integer Talbot effect undergone by the original unmodulated optical pulse train through the dispersive lines in the network was verified beforehand by using an optical sampling oscilloscope (EXFO, PSO-102).

Supplementary Information for

**Temporal optical neurons for serial deep learning**

# Temporal optical neurons for serial deep learning

Zhixing Lin,[1,2,3], Shuqian Sun [1,2,3], José Azaña, [4] Wei Li [1,2,3], Ninghua Zhu[1,2,3], and Ming Li,[1,2,3,*]

*To whom correspondence should be addressed; E-mail: ml@semi.ac.cn*

[1] State Key Laboratory on Integrated Optoelectronics, Institute of Semiconductors, Chinese Academy of Sciences, Beijing 100083, China
[2] School of Electronic, Electrical and Communication Engineering, University of Chinese Academy of Sciences, Beijing 100049, China
[3] Center of Materials Science and Optoelectronics Engineering, University of Chinese Academy of Sciences, Beijing 100190, China
[4] Institut National de la Recherche Scientifique – Centre Energie, Matériaux et Télécommunications (INRS-EMT), Montreal, Quebec, Canada.

## 1. Optical field analysis of serial ONN

### 1.1 Forward propagation

We begin this analysis at the input part of the serial ONN. The light source we use in this ONN strategy is a pulse train emitted by an actively mode-locked laser whose complex output field function can be written as:

$$E_{MML}(t) = \sum_i E_p(t - iT) e^{-j\varphi_0}, \tag{1}$$

where $T$ is the repetition period of the pulse train, $\varphi_0$ is the constant initial phase over time, and $E_p(t)$ is the expression of a single pulse:

$$E_p(t) = exp\left(\frac{-(t - T/2)^2}{2c_{lw}^2}\right), \tag{2}$$

where $c_{lw}$ is the parameter controlling the temporal linewidth of the pulse, such that $E_p(t) = 0$, when $|t| > T$. The object or waveform to be processed by the neural network ($S_{obj}(t)$) is sampled by this pulse train. This way, we obtain the input waveform to the serial ONN:

$$E_0(t) = E_{in}(t) = S_{obj}(t) \cdot \sum_i E_p(t - iT) e^{-j\varphi_0}, \tag{3}$$

For each layer of the serial ONN, the layer under consideration receives the output of the last layer ($E_{l-1}$), processes it and exports its processed waveform ($E_l$) to the next layer (the input of layer 1 is $E_0$). The processing contains two parts, as shown in Fig. 2a, i.e., temporal phase modulation and dispersion. The input of layer $l$ firstly experiences phase modulation, so we obtain:

$$E_{l1}(t) = E_{l-1}(t) \cdot exp(-j\varphi_l(t)), \qquad (4)$$

where $\varphi_l(t) = \sum_i \varphi_i^l \cdot ones(t - iT)$, is defined as a step function of which each period $T$ has one or more step heights $\varphi_i$, each confined within $2\pi$. All these step heights are the parameters to be trained. After the phase modulation, the wave will go through a dispersive medium with a specific dispersion value. For convenience, we model the operation of the dispersive medium in the frequency domain. In particular, the spectral transfer function of the dispersive medium can be expressed as follows:

$$H(\omega) = \exp(j|\ddot{\Phi}|\omega^2/2), \qquad (5)$$

where $|\ddot{\Phi}|$ is the first order dispersion coefficient and $\omega$ is the angular frequency. By multiplying the expression after phase modulation with the spectral transfer function and performing the inverse Fourier transform, we get the following expression for the waveform at the output of layer $l$:

$$\begin{aligned} E_l(t) &= \mathcal{F}^{-1}\{H(\omega) \cdot \mathcal{F}[E_{l1}(t)]\} \\ &= \mathcal{F}^{-1}\{H(\omega) \cdot \mathcal{F}[E_{l-1}(t) \cdot \varphi_l(t)]\}, \end{aligned}$$

(6)

The output of the serial ONN system is detected by a photodetector, which is only sensitive to optical intensity. Specifically, the intensity of the final output of the serial ONN, i.e., the output of layer L (assuming that this ONN has L layers), can be written as:

$$I_{out}(t) = |E_L(t)|^2. \qquad (7)$$

This is the profile that is used to calculate the loss function of the whole network in the backward propagation section of the system.

## 1.2 Backward propagation

The loss function is used as the indicator of the resemblance between the waveforms obtained at the output of the network and the target ideal outputs ("labels"). For this

purpose, the corresponding mean square error (MSE) is calculated:

$$e(\varphi_i^l) = \frac{1}{\Delta t}\int [I_{out}^{1/2}(t) - I_{label}^{1/2}(t)]^2 dt, \tag{8}$$

where $I_{label}(t)$ is the "label" output corresponding to a specific "feature" input. Because the independent variable $t$ is integrated, the MSE is independent of time and instead it depends on the trainable variable $\varphi_l(t)$ — the modulation phase at each layer. So the MSE will change as the phase of each layer — $\varphi_i^l$ — is modified.

In the back-propagation algorithm, the gradient $g = \nabla_{\varphi_i^l} e(\varphi_i^l)$ — the partial derivative between the MSE and the trainable variables — is calculated to update the variables. It determines the direction in which the modulation phase is adjusted in the backward propagation. Using this differential parameter, the modulation phase is consecutively updated to minimize the MSE as follows:

$$\varphi_{i,k}^l = \varphi_{i,k-1}^l - \alpha \cdot g_{k-1}, \tag{9}$$

where $k$ is the step of training, $\alpha$ is the learning rate, and $g_{k-1}$ is the gradient at step $k-1$. Beside the stochastic gradient rule as the update rule used in our models, there are many other rules that might be alternatively considered to perform this task.

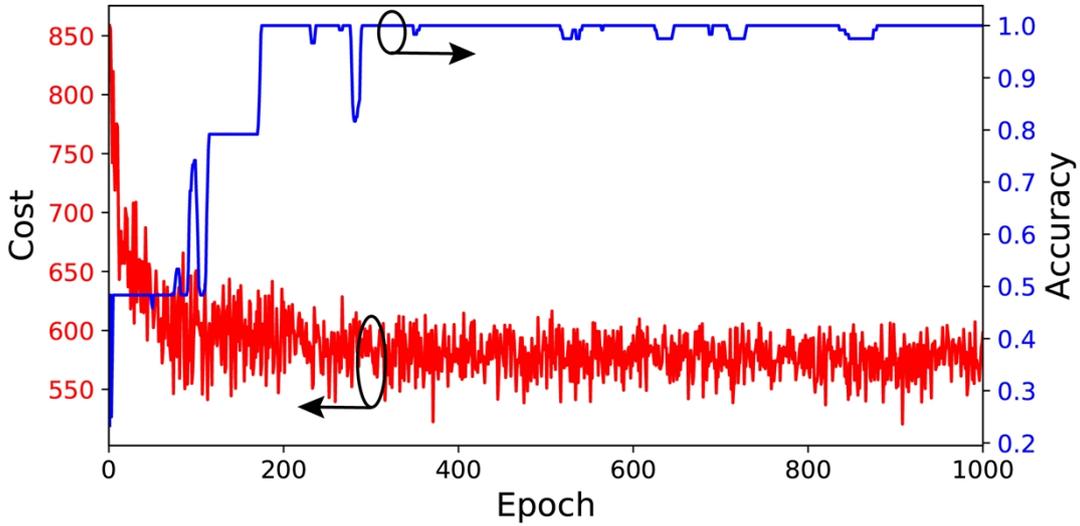

**Fig. S1|** The evolution of cost (red) and accuracy (blue) of the training process for the analog signals' classification problem considered in the main text.

## 2. Training details

### 2.1 Classification of analog signals

Concerning the training for the analog signals' classifier described in the main text, this was carried out with 1000 epochs, and the evolution profiles of the cost and accuracy functions during the training process are presented in Fig. S1. This figure confirms that cost function follows the desired stable evolution. The fluctuation of the accuracy trend in early epochs is due to the large initial learning rate, which is set to decay at a factor of 0.3 every 200 epochs. The ripple of the cost evolution relates to the sensibility of the final output of this serial ONN model to the phase of each layer. We anticipate that the fluctuations in the accuracy curve could be reduced using a smaller learning rate, but this would translate into an increased convergence time.

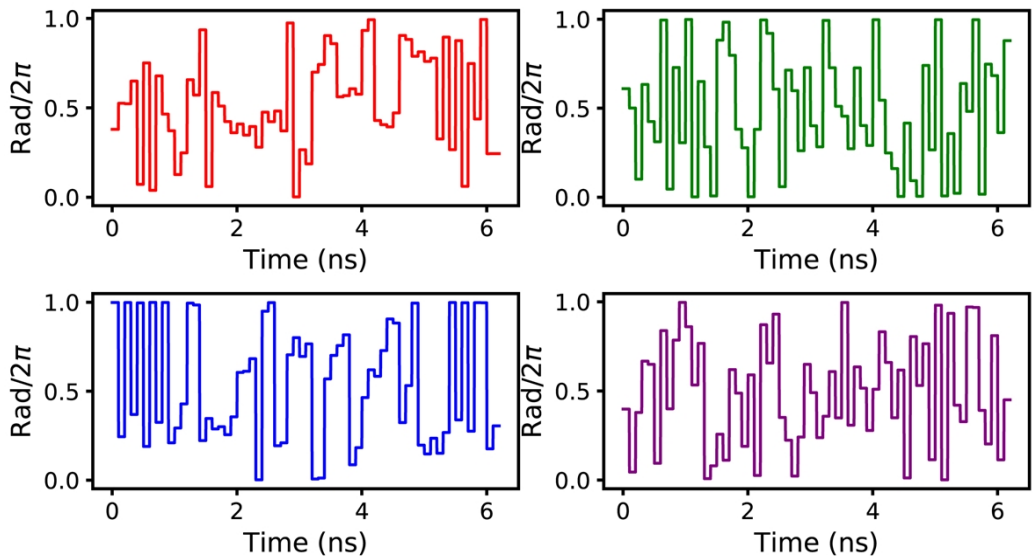

**Fig. S2|** The trained modulation phase profiles for layers 1, 2, 3 and 4 (red, green, blue, and purple lines, respectively) that are obtained for the considered analog signals' classification problem.

The trained modulation phase profiles of the four layers are shown in Fig. S2. They are restricted within the range between 0 and $2\pi$ through using a sigmoid function. Because the periodic pulses will shift half of the period over time by the integer temporal Talbot effect, the phase profile is shifted by half of the phase period to prevent the pulses from meeting with the phase jump points in the even-numbered layers. The data rate of the modulation phase is set to be twice the optical pulse repetition rate, i.e., 10 gigabits per second, which is still within the available range of a state-of-the-art electronic AWG (Keysight, M8194A, 120 GSa/s simultaneously for four channels[1]). A lower data rate of the phase profile would translate into the presence of an increased amount of noise around the main target peak in the output waveform, an issue that is further discussed in Supplementary Section 3.

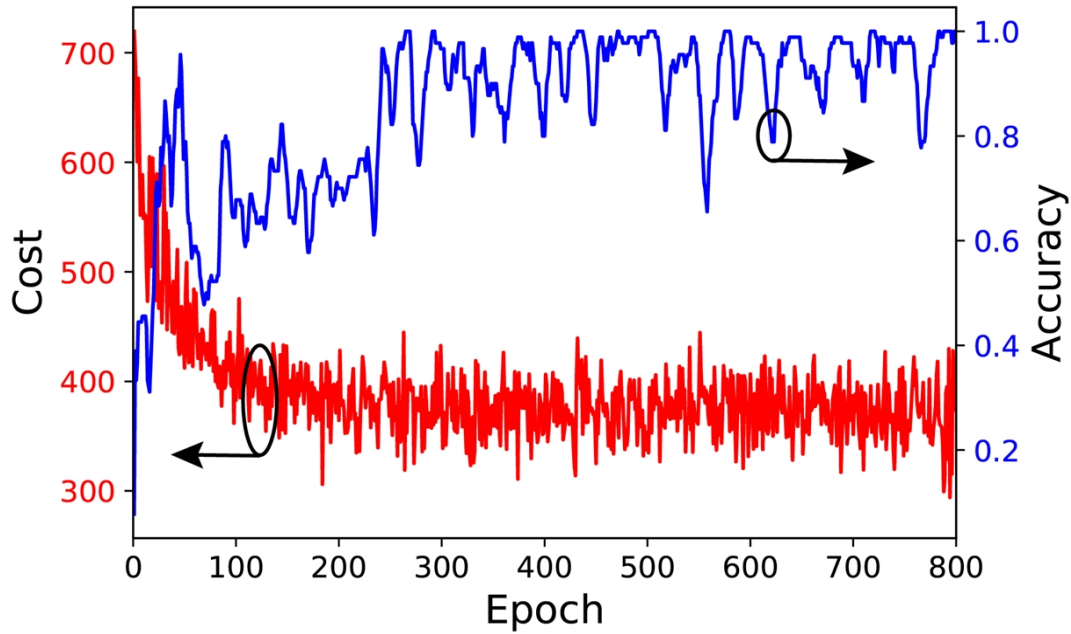

**Fig. S3|** The evolution of cost (red) and accuracy (blue) of the training process for the digital signals' classification problem considered in the main text.

## 2.2 Classification of digital signals

The evolution of the cost and accuracy functions, as well as the resulting modulation phase profiles, during the training process in the digital signals' classification problem considered in the main text are shown in Fig. S3 and Fig. S4, respectively. In Fig. S3, we can observe a significant fluctuation of the accuracy profile; we select an epoch of 800 as the training result by checking the outcome after every epoch such that to get a high success rate. Here, we select a training point for which the accuracy is not stable since the accuracy gets lower when it reaches an increased stability.

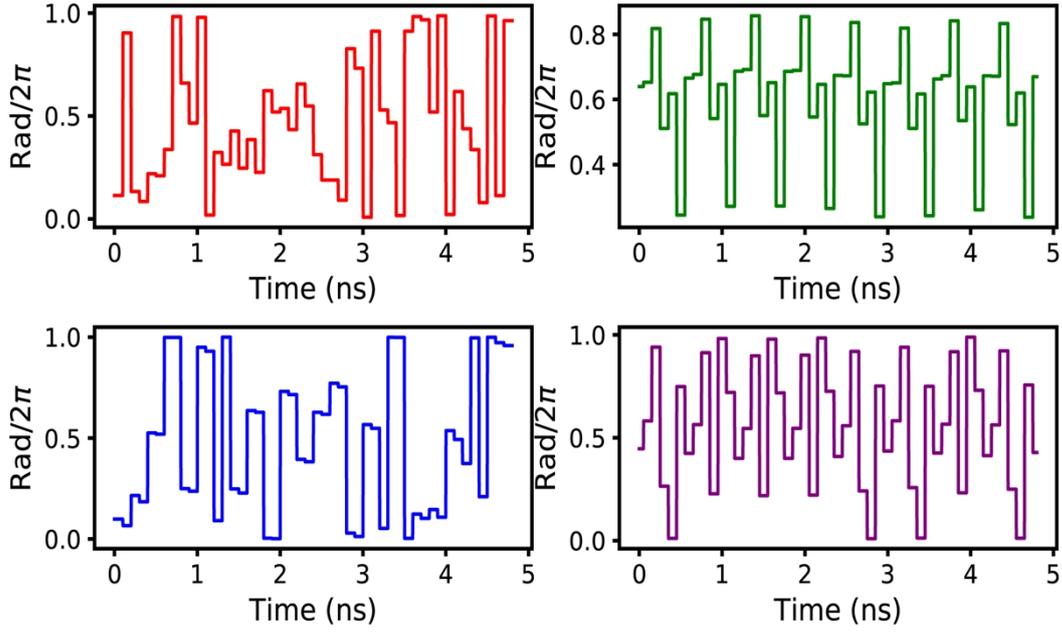

**Fig. S4|** The trained modulation phase profiles for layers 1, 2, 3 and 4 (red, green, blue, and purple lines, respectively) that are obtained for the considered digital signals' classification problem.

## 3. Considerations for the experimental configuration

As mentioned in the main text, the demonstrated experimental serial ONN platform was designed according to the equipment available to the team at the time of this work. Because there were only two AWG channels available in our labs, which could support the implementation of two layers of the network with phase modulation, we planned for the realization of a 2-layer serial ONN model. The limited depth of the model affects the capability of classification, such that this scheme can distinguish successfully 2 different patterns only. Moreover, the performance of the ONN with two phase-modulation layers only is still far from satisfactory, as shown in Fig. S5. In this figure, the trained classification results are shown in Fig. S5 a—f and Fig. S5 g—l for 'u' & 'c' and 'a' & 's', respectively. Not only the success rate is relatively low, say 84.4% for the classification of 'u' & 'c' and 64.4% for that of 'a' & 's', but additionally, the trained output waveforms lack the desired high contrast between the pulse used to mark the classification result and the neighboring pulses, as shown in Fig. S5 c, f, i and l. Fortunately, we find that an extra layer with no phase modulation can greatly improve the performance of the ONN. In particular, through this strategy, we achieve a far higher the contrast in the output waveforms, as shown in Fig. 4 in the main text, and the success rate is increased to 100%. This performance improvement is partly attributed to the fact that the layer with no phase modulation can be regarded as a layer with a constant phase modulation and therefore, this scheme actually implements a network with a larger depth, leading to the observed improvements in the feature extraction process.

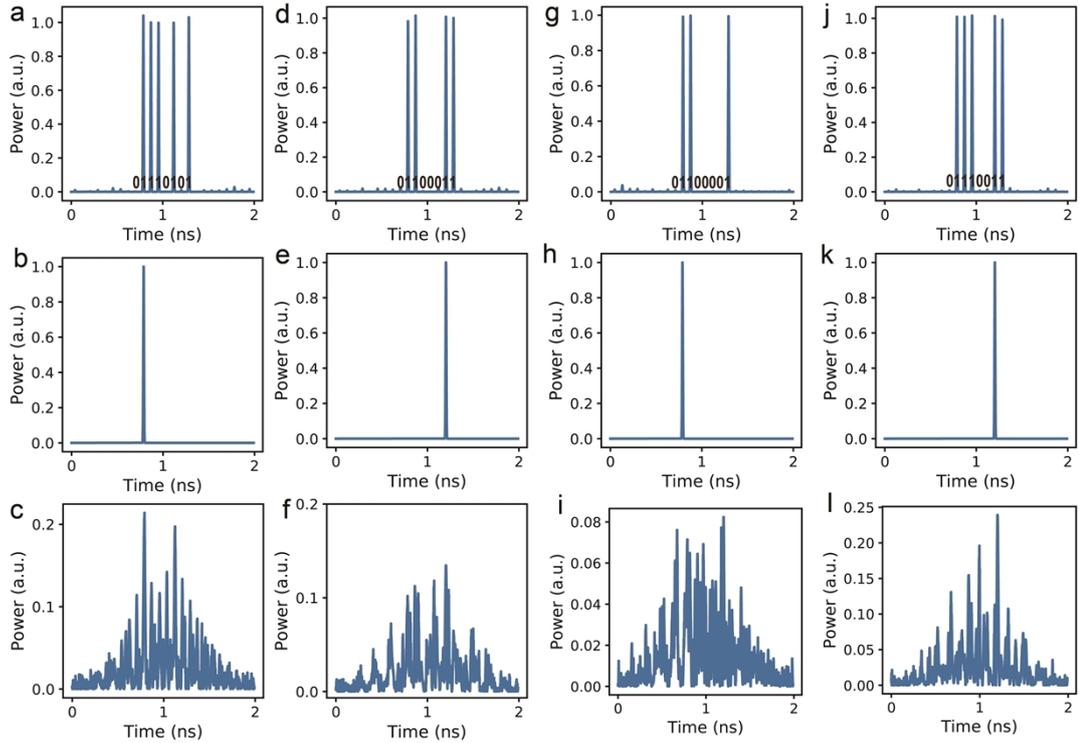

**Fig. S5| a, d, c, j,** 'Feature' waveforms of four English letters— 'u', 'c', 'c' and 's'—in the form of binary ASCII code. Each bit is represented by the power of each pulse. **b, e, h, k,** Ideal 'label' waveforms corresponding to **a, d, g,** and **j**. **c, f, i, l,** Trained output waveforms corresponding to **a, d, g,** and **j.**

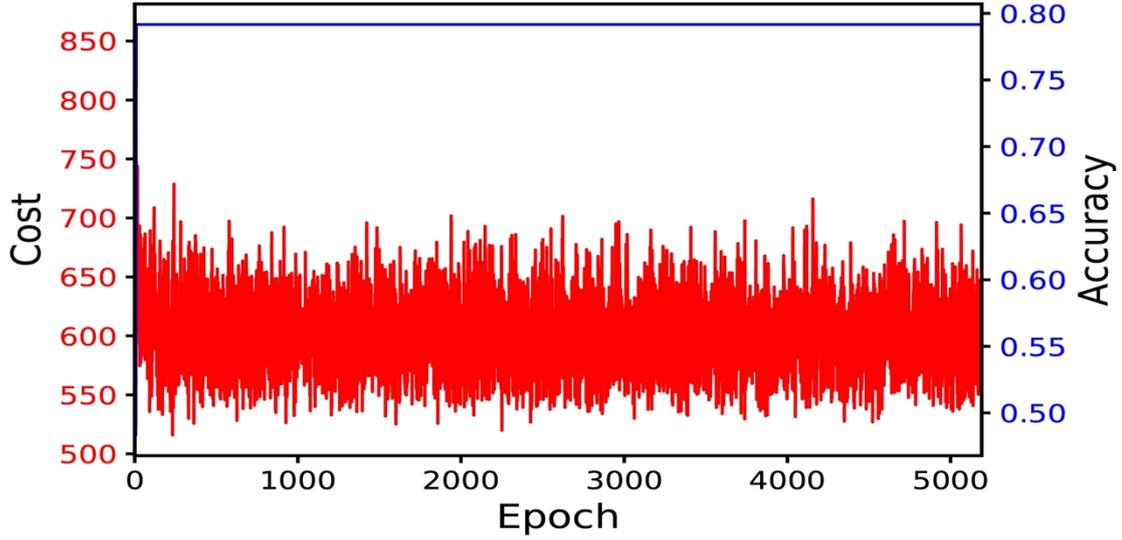

**Fig. S6|** The evolution of cost (red) and accuracy (blue) of the training process for the considered analog signals' classification problem when the dispersion of each layer is set to $2D_T + 200$ ps/nm ($D_T$ is the dispersion value satisfying the first-order integer temporal Talbot effect).

## 4. Selection of dispersion value

In regards to the specific dispersion value to be used at each layer, we have concluded that an optimal performance is achieved when choosing a dispersion that satisfies a temporal Talbot condition for the input optical pulse train. In general, the dispersion value at each layer will affect the performance (classification accuracy) of the serial ONN. As seen in Fig. S6, the accuracy of the considered deep learning system (for the analog signals' classification problem described in the main text) is significantly deteriorated when the applied dispersion does not satisfy a Talbot condition. In the example shown herein (with a deviation of 1.57% over the closest Talbot dispersion value), the accuracy can only reach a maximum of 79.2% even after many rounds of training (>5000). Also, we did several other simulations with different dispersion deviations, and got similar results. In sharp contrast, the accuracy of the ONN when the dispersion is set to satisfy a Talbot condition at every layer can reach 100%, as shown in Fig. S1. We attribute the improved performance obtained under Talbot conditions to the fact that operation under these conditions ensure a more uniform re-distribution of the energy of each dispersed pulse along the time domain, ensuring an optimal interaction (i.e., coherent addition) between consecutive temporal neurons at the positions of the corresponding temporal pulses.

# 5. Demonstration of classification for time-consecutive patterns (objects)

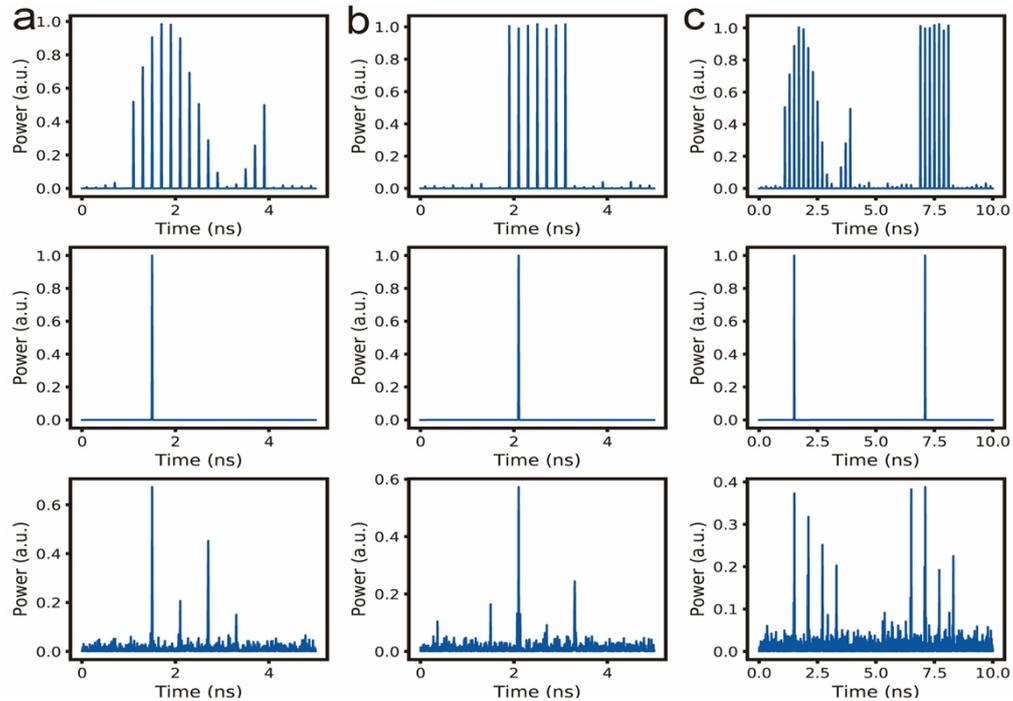

**Fig. S7|** The result of classification for single events and consecutive patterns using the same-signed dispersion of $D_T$ across the neural network. The first, second and third row represent the input patterns to be analyzed, the target ideal output waveforms and the simulated output waveforms, respectively.

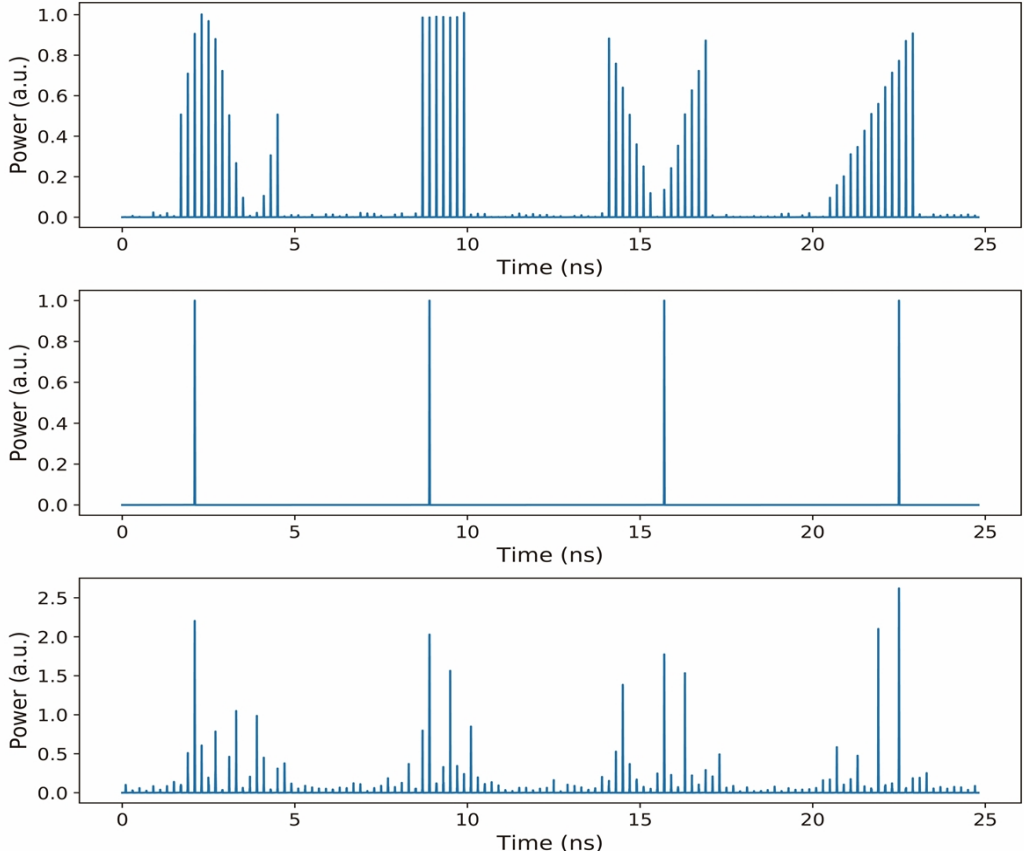

**Fig. S8|** The result of classification for consecutive patterns with dispersion of $+2D_T$, $-2D_T$, $+2D_T$ and $-2D_T$ when extending the gap between patterns. The first, second and third row represent input waveforms, target waveforms and output waveforms, respectively.

As mentioned in the main text, the proposed serial ONN is well suited for the analysis (e.g., classification) of different time-consecutive patterns or objects but in this case, the scheme needs to be carefully designed. A main consideration is that the dispersion in the system may induce undesired interference effects among the processed waveforms corresponding to different consecutive input patterns, negatively affecting the overall performance of the network. The trained phase profiles for a given input object should still provide the predicted performance when the same object is repeated along the time domain. However, a significant deterioration of the performance may be produced when different patterns are intended to be analyzed in a consecutive fashion, see results in Fig. S7. This is so because in this case, unintended interactions may be introduced between the waveform under analysis and adjacent waveforms that differ from those considered in the training process. In the example shown in Fig. S7c, the two different patterns in Figs. S7a and b are analyzed in a consecutive fashion with a gap in between them of 10× the input pulse repetition period (5 periods on each side) and assuming a dispersion in each layer of $D_T$. Clearly, the output waveform corresponding to each pattern deviates more significantly with respect to the ideal one than in the case of analysis of each of the patterns separately, Fig. S7 a and b, respectively. A more significant deterioration is still expected for a narrower gap or a higher dispersion. However, by properly choosing the amount of dispersion and the length of the gap in between consecutive input patterns, as well as using a symmetrical dispersion strategy (see below), one can ensure that the classification of consecutive patterns or objects is performed with the desired accuracy, as shown through the example in Fig. S8. In this latest example, the gap is set to 16× the pulse repetition period and a symmetrical dispersion strategy is utilized, in which the same amount of dispersion is used in consecutive layers but with an opposite sign. The pattern separation implies that an additional latency by the gap length (~3.2 ns in this example) should be considered for the analysis of each of the input patterns. The results in Fig. S8 show that the peak pulses of each output waveform are located in the designed temporal positions, confirming that the proposed serial ONN strategy can be designed for a successful analysis of different patterns as they arrive sequentially into the network.

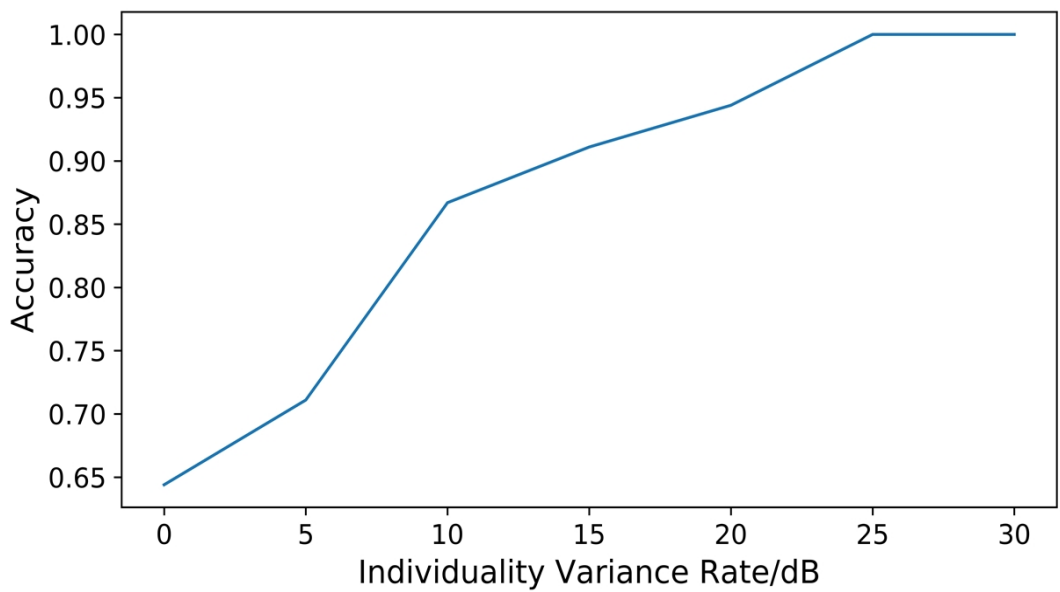

**Fig. S9|** Accuracy evolution of the serial deep learning scheme considered in the main text for the classification of digital signals, with IVRs from 0 to 30 dB.

# 6. Discussion about the influence of the IV to the accuracy of the digital classification

Intuitively, if the data used for training and test diversifies, i.e., if the individuals of the same kind of object differ from each other, it becomes harder for the neural network to perform an accurate classification of the incoming patterns. In conventional deep learning, this concept is called dissimilarity or similarity. Here, we evaluate this parameter using the individuality variance (IV), which can be quantified through the individuality variance rate (IVR):

$$\text{IVR} = 10\log_{10}\frac{P_s}{P_{IV}}, \tag{10}$$

where $P_s$ is the nominal peak amplitude of each pulse (after modulation by the pattern under analysis), and $P_{IV}$ quantifies the peak amplitude variance of each pulse. We used a Gaussian White Noise (GWN) function to generate the pulse-to-pulse amplitude difference, in which $P_{IV}$ is the standard deviation of the GWN. Then, the generated amplitude difference is imposed on each pulse amplitude. We train the SONN model for classification of the digital signals described in the main text, and we test the accuracy of the model at IVRs of 0 dB, 5 dB, 10 dB, 15 dB, 20 dB, 25dB, and 30 dB. The results of this study are shown in Fig. S9, confirming that the accuracy increases when the variance of the patterns/objects decreases. At an IVR of about 25 dB, the accuracy already reaches 100%. The reason why we choose the classification of digital patterns to explore the influence of the IVR to the SONN performance is that the IV added to the analog patterns will change the patterns greatly due to the different amplitudes between 0 and 1, which eliminates the features of the patterns. In contrast, the digital patterns are more capable of enduring the IV because the amplitude of the digital pattern is 1 or 0. Still, we explored the impact of the IV to the accuracy rate: the accuracy remains 79.2% or lower before the IVR reaches 30 dB.

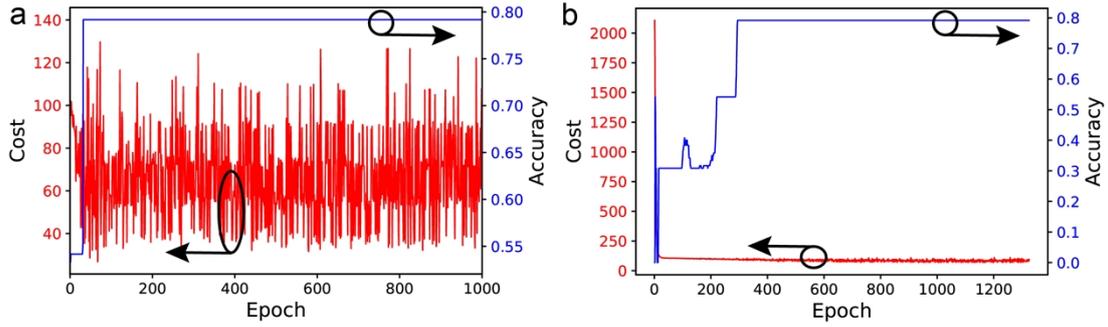

**Fig. S10| a,** The evolution of cost (red) and accuracy (blue) of the training process when the neurons are applied with pure amplitude modulation. **b,** The evolution of cost (red) and accuracy (blue) of the training process when the neurons are applied with complex modulation (phase and amplitude). The shown results are for the problem of analog signals' classification described in the main text, only differing in the modulation type applied to the neurons.

## 7. Comparison between the modulation types

In order to explore the influence of the modulation format to the recognition performance, we trained and simulated the SONNs with modulation types different to the phase-modulation case reported in the main text (amplitude modulation and complex modulation). Results are shown here for the analog signals' classification problem described in the main text (Fig. 2). Besides the modulation type, all other parameters remain identical to those defined for the problem at hand in the main text. The results of this extended analysis are shown in Fig. S10. These results show that the accuracy rates of both modulation types are 79.2% at best, which is a lower value than that obtained with using pure phase modulation. Considering that the rest of specifications are identical, except for the modulation type, the decrease of the recognition performance can be attributed to the change in the modulation type. Besides, numerous additional simulated trainings have been conducted to verify the reliability of this conclusion. However, we have also observed that the use of amplitude modulation helps in that in provides a significant mitigation of the undesired pulses around the main target peak pulse in the outcome waveform. If a recognition output with a high extinction ratio is preferred, the amplitude modulation could be added in to the SONN. Notice that although the extinction ratio is increased in this case, the absolute value of amplitude of the main peak is considerably decreased, i.e., approximately by 10 dB, which explains the observed deterioration in the obtained accuracy parameter.